\documentclass{jnmp}
\usepackage[intlimits]{amsmath}
\usepackage{graphicx}

\numberwithin{equation}{section}
\theoremstyle{definition}
\newtheorem*{definition}{Definition}

\renewcommand{\ba}{\begin{array}}
\renewcommand{\ea}{\end{array}}
\newcommand{\beg}{\begin{eqnarray}}
\newcommand{\eeq}{\end{eqnarray}}
\newcommand{\bg}{\begin{eqnarray*}}

\newcommand{\ed}{\end{eqnarray*}}

\newcommand{\td}{\tilde}

\newcommand{\bigint}{\mbox{\LARGE
$\displaystyle{\int}$}}

\newcommand{\I}{{ i}}
\newcommand{\XI}{{x_{ i}}}
\newcommand{\TI}{{t_{ i}}}
\newcommand{\J}{{ j}}
\newcommand{\XJ}{ {x_{ j}}}
\newcommand{\TJ}{ {t_{ j}}}

\newcommand{\XN}{ {x_{ 0}}}
\newcommand{\TN}{ {t_{ 0}}}

\newcommand{\one}{{u}_{ 1}}
\newcommand{\tone}{{\tilde u}_{ 1}}
\newcommand{\XO}{ {x_{ 1}}}
\newcommand{\TO}{ {t_{ 1}}}
\newcommand{\tXO}{ {\tilde x_{ 1}}}
\newcommand{\tTO}{ {\tilde t_{ 1}}}

\newcommand{\two}{{u}_{ 2}}
\newcommand{\ttwo}{{\tilde u}_{ 2}}
\newcommand{\XTW}{ {x_{ 2}}}
\newcommand{\TTW}{ {t_{ 2}}}
\newcommand{\tXTW}{ {\tilde x_{ 2}}}
\newcommand{\tTTW}{{\tilde t_{ 2}}}

\newcommand{\three}{ {u}_{ 3}}
\newcommand{\tthree}{ {\tilde u}_{ 3}}
\newcommand{\XTH}{ {x_{ 3}}}
\newcommand{\TTH}{ {t_{ 3}}}
\newcommand{\tXTH}{ {\tilde x_{ 3}}}
\newcommand{\tTTH}{ {\tilde t_{ 3}}}

\newcommand{\tfour}{ {\tilde u}_{ 4}}
\newcommand{\XF}{ {x_{ 4}}}
\newcommand{\TF}{ {t_{ 4}}} 
\newcommand{\tXF}{ {\tilde x_{ 4}}}
\newcommand{\tTF}{ {\tilde t_{ 4}}}

\newcommand{\five}{ {u}_{ 5}} 
 
\newcommand{\XV}{ {x_{ 5}}} 
\newcommand{\TV}{ {t_{ 5}}} 
\newcommand{\tXV}{ {\tilde x_{ 5}}} 
\newcommand{\tTV}{ {\tilde t_{ 5}}}

\newcommand{\tsix}{ {\tilde u}_{ 6}} 
\newcommand{\XS}{ {x_{ 6}}} 
\newcommand{\TS}{ {t_{ 6}}} 
\newcommand{\tXS}{ {\tilde x_{ 6}}} 
\newcommand{\tTS}{ {\tilde t_{ 6}}}

\newcommand{\seven}{ {u}_{ 7}} 
\newcommand{\tseven}{ {\tilde u}_{ 7}} 
\newcommand{\XSV}{ {x_{ 7}}} 
\newcommand{\TSV}{ {t_{ 7}}} 
\newcommand{\tXSV}{ {\tilde x_{ 7}}} 
\newcommand{\tTSV}{ {\tilde t_{ 7}}}

\newcommand{\teight}{{\tilde u}_{ 8}} 
\newcommand{\XE}{ {x_{ 8}}} 
\newcommand{\TE}{ {t_{ 8}}} 
\newcommand{\tXE}{ {\tilde x_{ 8}}} 
\newcommand{\tTE}{ {\tilde t_{ 8}}}

\renewcommand{\p}{\partial} 
 
\newcommand{\notlhd}{\lhd\kern-.8em{/}\ } 
\newcommand{\notexist}{\ \exists\kern-.5em{\raise.1em\hbox{/}}\ }

\newcommand{\pde}[2]{\frac{\p #1}{\p #2}} 
\newcommand{\ppd}[3]{\frac{\p^2 #1}{\p #2\p #3}} 
\newcommand{\pdd}[2]{\frac{\p^2 #1}{\p #2^2}} 
\newcommand{\inp}{{\mbox{\vbox{\hrule width0ex\hbox{\vrule
 height0ex\kern3.8pt
\vbox{\kern2.5pt}\kern3.8pt \vrule height1.6ex}
\hrule width1.6ex}}}}

\begin{document}

\renewcommand{\evenhead}{N Euler and M Euler}
\renewcommand{\oddhead}{Linearisable second-order evolution equations}


\thispagestyle{empty}

\begin{flushleft}
\footnotesize \sf
\end{flushleft}

\FirstPageHead{8}{3}{2001}
{\pageref{euler-firstpage}--\pageref{euler-lastpage}}{Article}


\copyrightnote{2001}{N Euler and M Euler }

\Name{A Tree of Linearisable Second-Order Evolution Equations
by Generalised Hodograph Transformations  }

\label{euler-firstpage}

\Author{Norbert EULER and Marianna EULER}

\Address{Department of Mathematics,  Lule\aa\ University of Technology, \\
SE-971 87 Lule\aa, Sweden\\
E-mails: Norbert@sm.luth.se, Marianna@sm.luth.se}

\Date{Received\,\footnote{Communicated by P G L Leach}
October 23, 2000; Revised March 14, 2001, Accepted
April 26, 2001}

\strut\hfill

\strut\hfill{\small{\sc To the memory of \qquad\qquad}}

\strut\hfill{\small{\sc Wilhelm Fushchych \qquad}}

\strut\hfill

\begin{abstract}
\noindent
We present a list of $(1+1)$-dimensional
second-order evolution equations all connected via a
proposed generalised hodograph transformation, resulting in a tree of 
equations transformable to the linear second-order autonomous evolution 
equation. The list includes autonomous and 
nonautonomous equations.
\end{abstract}



\section{Introduction} 

In \cite{eul} we report on the linearisation of the 
hierarchy of evolution equations
\begin{gather}
\label{hier}
u_t=R^m[u](u^{-2}u_x)_x,\qquad R[u]=D_x^2u^{-1}D_x^{-1}, \qquad m=0,1,2,\ldots
\end{gather}
by an extended hodograph transformation and define an autohodograph
transformation for this hierarchy. The autohodograph transformation
is revealed by the composition of the extended hodograph
transformation and the linearising contact transformation.
The extended hodograph
transformation for the case $m=0$, first introduced in \cite{cfa},
is of the form
\begin{gather}
dX(x,t)=udx+\left(\int^x u_t(\xi,t)d\xi\right) dt\notag\\
dT(x,t)=dt\\[3mm]
U(X,T)=x.\notag
\end{gather}

In the present paper we generalise the extended hodograph
transformation and name it an {\it $x$-generalised hodograph
transformation}. 
We are interested to derive a class (or {\it tree}, as we
prefer to call it) of $(1+1)$-dimensional second-order
evolution equations which are linearisable. This tree of equations,
containing arbitrary nonconstant functions in $C^2(\Re)$, 
is constructed by nonlinearising the general second-order linear
autonomous equation using the $x$-generalised hodograph transformation.
In this way both nonlinear autonomous and nonlinear nonautonomous
equations are revealed. The linearising transformations
are 
obtained by composing and inverting the appropriate $x$-generalised
hodograph transformations. Besides the obvious examples, such as
Burgers' equation and (\ref{hier}) (with $m=0$),
our tree of equations consists of new linearisable equations
as well as several cases found in 
the literature (for example \cite{cfa, cal, sok, Zak, fok, is, fy}).
In particular, the results obtained by Sokolov, Svinolupov
and Wolf \cite{sok} are
special cases of
our tree of equations.


The paper is organized as follows: In Section 2 we define the
$x$-generalised hodograph transformation and introduce the 
notation.
The most general form of the second-order equation linearisable
by the proposed method is established in this section.
In Section 3 we consider
autonomous second-order evolution equations and derive a tree
of linearisable equations. The linearising transformations
are listed explicitly.
The $x$-generalised hodograph transformations generating the equations
are
given in the Appendix. Only one equation from the tree
of equations
admits an autohodograph transformation in the sense of \cite{eul}.
It should be pointed out
that
under the proposed $x$-generalised hodograph transformation
the tree of autonomous
linearisable equations (see Diagram 1) is complete. Some examples are
given.
In Section 4 we list
nonautonomous
linearisable second-order evolution
equations which are generated from the tree of autonomous 
linearisable equations (Diagram 1) by $x$-generalised hodograph
transformations.
This case is not complete 
as we consider only the case where
the coefficient of the highest derivative is autonomous.
Once again we give the linearising transformations explicitly
as well as some examples.
The corresponding $x$-generalised hodograph transformations are listed in
the Appendix.
In the nonautonomous case each linearisable equation contains two arbitrary
functions
in $C^2(\Re)$; one function depending on the dependent variable and
one depending on the independent ``space''-variable $x$.


\section{The $x$-Generalised hodograph transformation}

\begin{definition}
{
The transformation
\beg
\label{2nd_hodo}
_n\mbox{\bf H}_j^i:\left\{\ba{l}
\displaystyle{d\XI(\XJ,\TJ )
=f_1(\XJ ,u_\J)dx_\J+f_2(x_\J,u_\J,u_{\J\XJ}, u_{\J\XJ\XJ},\ldots
, u_{\J{x_j^{n-1}}}
)d\TJ}
\\[3mm]
d\TI(\XJ,\TJ)=d\TJ \\[3mm]
u_\I(\XI,\TI )=g(\XJ),
\ea\right.
\eeq
with $i\neq j$, $n=2,3,\ldots$ and
\beg
\label{poincare}
u_{\J\TJ}\pde{f_1 }{u_\J}=
\pde{f_2 }{\XJ}+u_{\J\XJ}\pde{f_2 }{u_\J}+
u_{\J\XJ\XJ}\pde{f_2 }{u_{\J\XJ}}
+\cdots +u_{\J{x_j^{n}}}\pde{f_2}{u_{\J{x_j^{n-1}}}},
\eeq
is called an $x$-generalised hodograph transformation.
}
\end{definition}

\strut\hfill

\noindent
{\bf Remarks:} We named the above transformation $x$-generalised
in order to have the possibility in future to introduce other
generalisations of the extended hodograph transformation. Condition
(\ref{poincare}) follows from the Lemma of Poincar\'e, i.e.,
$d(dx_i)\equiv 0$.

\strut\hfill

\noindent
Here and below the subscripts denote partial derivatives, e.g.
\bg
u_{j\XJ\XJ}=\pdd{u_j}{\XJ}.
\ed

Consider a general $(1+1)$-dimensional
second-order autonomous evolution equation
with dependent variable $u_i$ and independent variables $x_i,\ t_i$, viz.
\beg
\label{gen_F}
u_{i\TI}=F(u_i, u_{i\XI},u_{i\XI\XI}).
\eeq
Applying the $x$-generalised hodograph transformation (\ref{2nd_hodo})
leads to the following particular form for $f_2$:
\beg
\label{gen_f2}
f_2(x_j,u_j,u_{j\XJ})=-\frac{f_1(\XJ,u_j)}{\dot g(\XJ)}
\left.\left[\vphantom{\frac{f_1(\XJ,u_j)}{\dot g(\XJ)}}
F(u_i,u_{i\XI},u_{i\XI\XI})\right]\right|_{\Omega}\ ,
\eeq
where
\begin{gather}
\label{omega}
\Omega=\left\{u_i=g(\XJ),\ u_{i\XI}
=\frac{\dot g(\XJ)}{f_1(\XJ,u_j)}\right.,
\notag\\[3mm]
 \qquad\qquad\left.u_{i\XI\XI}=\frac{\ddot g(\XJ)}{f_1^2(\XJ,u_j)}
-\frac{\dot g(\XJ)}{f_1^3(\XJ,u_j)}
\left(\pde{f_1}{\XJ}+\pde{f_1}{u_j}u_{j\XJ}\right)\right\}\ .
\end{gather}
The most general equation which results when transforming
(\ref{gen_F}) by the $x$-generalised hodograph transformation
(\ref{2nd_hodo}) with (\ref{gen_f2}) is
\begin{gather}
\label{long}
\pde{f_1}{u_j}u_{j\TJ}=\left[
\frac{1}{f_1^2}\left(\pde{f_1}{u_j}u_{j\XJ\XJ}
+\pdd{f_1}{u_j}u^2_{j\XJ}
+2\ppd{f_1}{\XJ}{u_j} u_{jx_j}
+\pdd{f_1}{\XJ}\right)\right.\notag\\[3mm]
\qquad\qquad-\frac{3}{f_1^3}\left(\pde{f_1}{\XJ}
+\pde{f_1}{u_j}u_{j\XJ}\right)^2
+\frac{3\ddot g}{\dot g f_1^2}
\left(\pde{f_1}{\XJ}+\pde{f_1}{u_j}u_{j\XJ}\right)
\left.\left.
-\frac{\dddot g }{\dot g f_1}\right]
\left[\pde{F}{u_{i\XI\XI}}\right]\right|_{\Omega}\notag\\[3mm]
\qquad\qquad\left.+\left[\frac{1}{f_1}
\left(\pde{f_1}{\XJ}+\pde{f_1}{u_j}u_{j\XJ}\right)
-\frac{\ddot g}{\dot g}\right]
\left[\pde{F}{u_{i\XI}}\right]\right|_{\Omega}
-f_1\left.\left[\pde{F}{u_{i}}\right]\right|_{\Omega}\notag\\[3mm]
\qquad\qquad-\left.\left[\frac{1}{\dot g}
\left(\pde{f_1}{\XJ}+\pde{f_1}{u_j}u_{j\XJ}\right)
-\frac{\ddot g f_1}{\dot g^2}\right]\left[\vphantom{\pde{f_1}{\XJ}}
F\right]\right|_{\Omega}\ .
\end{gather}
Here and below $\dot g$ denotes the derivative w.r.t. $x_j$, 
$\ddot g$ the second derivative w.r.t. $x_j$, etc.

Using (2.5) it can easily be
shown that the most general equation
which may be constructed by applying (\ref{2nd_hodo}) to the linear equation
\beg
\label{lin_auto}
u_{\I\TI}=u_{\I\XI\XI}+\lambda_1 u_{\I\XI}
+\lambda_2 u_\I,\qquad \lambda_1,\lambda_2\in
\Re
\eeq
is of the form
\beg
\label{gen-nonauto}
u_{\J\TJ}=F_1(\XJ, u_\J)u_{\XJ\XJ}
+F_2(\XJ, u_\J)u_{\XJ}
+F_3(\XJ, u_\J)u_{\XJ}^2
+F_4(\XJ, u_\J)
\eeq
for all iterations of the $x$-generalised hodograph transformation. The following
statment is therefore true:

\strut\hfill

\noindent
{\bf Proposition:}
{\it The most general $(1+1)$-dimensional
second-order evolution equation which may be
constructed to be linearisable in (\ref{lin_auto}) by the $x$-generalised
hodograph transformation (\ref{2nd_hodo})
is necessarily of the form (\ref{gen-nonauto}).}

\strut\hfill


\noindent
{\bf Remark:} In the sense of \cite{eul} an $x$-generalised hodograph
transformation which keeps an equation invariant is known as
an autohodograph transformation.

\strut\hfill

Finally we introduce an important notation which we use
throughout this paper in order to abbreviate the derivatives of
some arbitrary functions that appear in our tree of equations:
Let $f=f(\xi)\in C^2(\Re)$ with $df/d\xi\neq 0$.

Then we define the following bracket:
\beg
\label{bracket}
\left\{f\right\}_{\xi}:=-\frac{1}{2}\frac{df}{d\xi}
+f\left(\frac{df}{d\xi}\right)^{-1}\frac{d^2 f}{d\xi^2}.
\eeq

\section{Linearisable autonomous second-order equations}

\noindent
Here we give the
 second-order linearisable autonomous evolution equations constructed by
(\ref{2nd_hodo}). We found eight cases, listed below, resulting in a
 tree of equations shown in Diagram 1. By nonlinearising
(\ref{lin_auto}) with (\ref{2nd_hodo}) and restricting ourselves to
autonomous equations, we obtain Cases I, II, III, V and VII. These
equations follow when (\ref{2nd_hodo}) is applied to each
resulting autonomous equation until the iteration stops. That is,
until no new autonomous equation appears.
This happens at
eq. (\ref{int_3}), i.e., Case III. Applying the same procedure
but starting from the second-order semilinear equation
\bg
u_{i\TI}=u_{i\XI\XI}+G(u_i,u_{i\XI})
\ed
results in Cases IV, VI, and VIII. The corresponding linearising
transformations, given below for each case, are obtained by composing
and inverting the appropriate $x$-generalised hodograph transformations, given in the
Appendix.

\vspace{4mm}

\noindent
{\bf Case I:} Let $\lambda_1\in \Re$ and $h_1\in C^2(\Re)$
with $dh_1/du_1\neq 0$. Then
\beg
\label{int_1}
u_{1\TO}=h_1(u_1)u_{1\XO\XO}+\left\{h_1\right\}_{\one} u^2_{1\XO}
\eeq
is linearised to $u_{0\TN}=u_{0\XN\XN}+\lambda_1 u_{0\XN}$ by the
transformation
\bg
\label{lin_trans_1}
_2\mbox{\bf L}_0^1:\left\{\ba{l}
\displaystyle{x_1(x_0,t_0)=u_0}
\\[3mm]
dt_1(x_0,t_0)=dt_0\\[3mm]
\displaystyle{h_1(u_1(x_1,t_1))=u_{0\XN}^2. }
\ea\right.
\ed

\vspace{4mm}

\noindent
{\bf Case II:} Let $\lambda\in \Re\backslash\{0\}$,
$\lambda_1\in \Re$ and $h_2\in C^2(\Re)$ with $dh_2/du_2\neq 0$.
Then
\beg
\label{int_2}
u_{2\TTW}=h_2(u_2)u_{2\XTW\XTW}+\lambda h_2(u_2)u_{2\XTW}+
\left\{h_2\right\}_{\two} u^2_{2\XTW}
\eeq
is linearised to $u_{0\TN}=u_{0\XN\XN}+\lambda_1 u_{0\XN}$ by the
transformation
\bg
\label{lin_trans_2}
_2\mbox{\bf L}_0^2:\left\{\ba{l}
\displaystyle{x_2(x_0,t_0)=\frac{1}{\lambda}\ln|u_{0\XN}|
}
\\[3mm]
dt_2(x_0,t_0)=dt_0\\[3mm]
\displaystyle{h_2(u_2(x_2,t_2))=\frac{1}{\lambda^2}
\left(
\frac{u_{0\XN\XN}}{
u_{0\XN}}\right)^2.
}
\ea\right.
\ed

\vspace{4mm}

\noindent
{\bf Case III:} Let $\lambda_1\in \Re$, $\lambda_2\in
\Re\backslash\{0\}$ and $h_3\in C^2(\Re)$ with $dh_3/du_3\neq 0$.
Then
\beg
\label{int_3}
u_{3\TTH}=h_3(u_3)u_{3\XTH\XTH}+
\left\{h_3\right\}_{\three} u^2_{3\XTH}+2\lambda_2h_3^{3/2}(u_3)
\left(\frac{dh_3}{du_3}\right)^{-1}
\eeq
is linearised to $u_{0\TN}=u_{0\XN\XN}+\lambda_1 u_{0\XN}$
by the
transformation
\bg
\label{lin_trans_3}
_2\mbox{\bf L}_0^3:\left\{\ba{l}
\displaystyle{x_3(x_0,t_0)=\frac{2}{\lambda_2}\left(
\frac{u_{0\XN\XN}}{
u_{0\XN}}\right)
}
\\[3mm]
dt_3(x_0,t_0)=dt_0\\[3mm]
\displaystyle{h_3(u_3(x_3,t_3))=\frac{4}{\lambda_2^2}\left[\frac{\p}{\p \XN}
\left(\frac{u_{0\XN\XN}}{u_{0\XN}}\right)\right]^2.
}
\ea\right.
\ed

\vspace{4mm}

\noindent
{\bf Case IV.1:} Let $\{\lambda_1,\lambda_4\}\in \Re$,
$\{\lambda,\lambda_2\}\in\Re\backslash\{0\}$ and
$h_4\in C^1(\Re)\backslash \{0\}$.
Then
\beg
\label{int_4}
u_{4\TF}=u_{4\XF\XF}+\lambda_4 u_{4\XF}+\frac{1}{h_4(u_4)}
\left(\lambda_2-\frac{dh_4}{du_4}\right)u^2_{4\XF}+h_4(u_4)
\eeq
is linearised to $u_{0\TN}=u_{0\XN\XN}+\lambda_1 u_{0\XN} +\lambda_2 u_0$ by the
transformation
\bg
\label{lin_trans_4}
_2\mbox{\bf L}_0^{4.1}:\left\{\ba{l}
\displaystyle{dx_4(x_0,t_0)=
dx_0
+(\lambda_1-\lambda_4)d\TN
}
\\[3mm]
dt_4(x_0,t_0)=dt_0\\[3mm]
\displaystyle{\int^{u_4(x_4,t_4)}\frac{d\xi}{h_4(\xi)}=\frac{1}{\lambda_2}\ln
\left|\lambda u_0(x_0,t_0)\right|. }
\ea\right.
\ed

\vspace{4mm}

\noindent
{\bf Case IV.2:} Let $\lambda_1\in \Re$,
$\lambda_3\in\Re\backslash\{0\}$, $\lambda_4\in \Re$ and
$h_4\in C^1(\Re)\backslash \{0\}$.
Then
\beg
\label{int_42}
u_{4\TF}=u_{4\XF\XF}+\lambda_4 u_{4\XF}-\frac{1}{h_4(u_4)}
\frac{dh_4}{du_4}u^2_{4\XF}+h_4(u_4)
\eeq
is linearised to $u_{0\TN}=u_{0\XN\XN}+\lambda_1 u_{0\XN}$ by the
transformation
\bg
\label{lin_trans_42}
_2\mbox{\bf L}_0^{4.2}:\left\{\ba{l}
\displaystyle{dx_4(x_0,t_0)=
dx_0
+(\lambda_1-\lambda_4)d\TN
}
\\[3mm]
dt_4(x_0,t_0)=dt_0\\[3mm]
\displaystyle{
\frac{1}{h_4(u_4(\XF,\TF))}\pde{u_4}{\XF}=-\frac{u_0(\XN,\TN)}{\lambda_3}.
}
\ea\right.
\ed

\vspace{4mm}

\noindent
{\bf Case V:} Let $\{\lambda,\lambda_2\}\in\Re\backslash\{0\}$,
$\lambda_1\in \Re$ and $h_5\in C^2(\Re)$ with $dh_5/du_5\neq 0$.
Then
\beg
\label{int_5}
u_{5\TV}=h_5(u_5)u_{5\XV\XV}
+\left(\lambda h_5(u_5)-\frac{\lambda_2}{\lambda}\right)u_{5\XV}
+ \left\{h_5\right\}_{\five} u^2_{5\XV}
\eeq
is linearised to $u_{0\TN}=u_{0\XN\XN}+\lambda_1 u_{0\XN} +\lambda_2 u_0$ by the
transformation
\bg
\label{lin_trans_5}
_2\mbox{\bf L}_0^5:\left\{\ba{l}
\displaystyle{x_5(x_0,t_0)=\frac{1}{\lambda}\ln|\lambda u_0|}
\\[3mm]
dt_5(x_0,t_0)=dt_0\\[3mm]
\displaystyle{h_5(u_5(x_5,t_5))=\frac{1}{\lambda^2}
\left(\frac{u_{0\XN}}{u_0}\right)^2
}
\ea\right.
\ed

\vspace{4mm}

\noindent
{\bf Case VI:} Let $\lambda_1\in \Re$ and $h_6\in C^2(\Re)$
with $dh_6/du_6\neq 0$.
Then
\beg
\label{int_6}
u_{6\TS}=u_{6\XS\XS}+h_6(u_6)u_{6\XS}
+\frac{d^2h_6}{du_6^2}\left(\frac{dh_6}{du_6}\right)^{-1}u^2_{6\XS}
\eeq
is linearised to $u_{0\TN}=u_{0\XN\XN}+\lambda_1 u_{0\XN}$ by the
transformation
\bg
\label{lin_trans_6}
_2\mbox{\bf L}_0^6:\left\{\ba{l}
\displaystyle{dx_6(x_0,t_0)=d\XN+\lambda_1 d\TN  }
\\[3mm]
dt_6(x_0,t_0)=dt_0\\[3mm]
\displaystyle{h_6(u_6(x_6,t_6))=2\left(\frac{u_{0\XN\XN}}{u_{0\XN}}\right). }
\ea\right.
\ed

\vspace{4mm}

\noindent
{\bf Case VII:} Let $\lambda_1\in \Re$,
$\lambda_3\in\Re\backslash\{0\}$
and $h_7\in C^2(\Re)$ with $dh_7/du_7\neq 0$.
Then
\beg
\label{int_7}
u_{7\TSV}=h_7(u_7)u_{7\XSV\XSV}+\lambda_3 u_{7\XSV}+
\left\{h_7\right\}_{\seven} u^2_{7\XSV}
\eeq
is linearised to $u_{0\TN}=u_{0\XN\XN}+\lambda_1 u_{0\XN}$ by the
transformation
\bg
\label{lin_trans_7}
_2\mbox{\bf L}_0^7:\left\{\ba{l}
\displaystyle{dx_7(x_0,t_0)=u_0d\XN+\left(u_{0\XN}+\lambda_1
u_0-\lambda_3
\right)d\TN}
\\[3mm]
dt_7(x_0,t_0)=dt_0\\[3mm]
\displaystyle{h_7(u_7(x_7,t_7))=u_0^2. }
\ea\right.
\ed

\vspace{4mm}

\noindent
{\bf Case VIII:}  Let $\lambda\in \Re\backslash\{0\}$,
$\{\lambda_1,\lambda_8\}\in \Re$
and $h_8\in C^2(\Re)$.  Then
\beg
\label{int_8}
& & u_{8\TE}=u_{8\XE\XE}+\lambda_8 u_{8\XE}+
h_8(u_8) u^2_{8\XE}
\eeq
is linearised to $u_{\TN}=u_{0\XN\XN}+\lambda_1 u_{0\XN}$ by the
transformation
\bg
_2\mbox{\bf L}_0^8:\left\{\ba{l}
\displaystyle{dx_8(x_0,t_0)=d\XN+(\lambda_1-\lambda_8)d\TN}\\[3mm]
dt_8(x_0,t_0)=dt_0\\[3mm]
\ba[b]{c}
\scriptstyle u_8(x_8,t_8)\\
\bigint
\ea
\displaystyle{
\left\{\exp\left(\int^\xi
h_8(\eta)d\eta\right) \left[\lambda\int^\xi\exp\left(\int^{\eta}
h_8(\eta')d\eta'\right)d\eta\right]^{-1}\right\}d\xi
}\\[3mm]
\displaystyle{\qquad =\frac{1}{2\lambda}\ln\left( u_{0\XN}^2\right).}
\ea\right.
\ed

\begin{figure}
\begin{center}
\includegraphics[width=16cm]{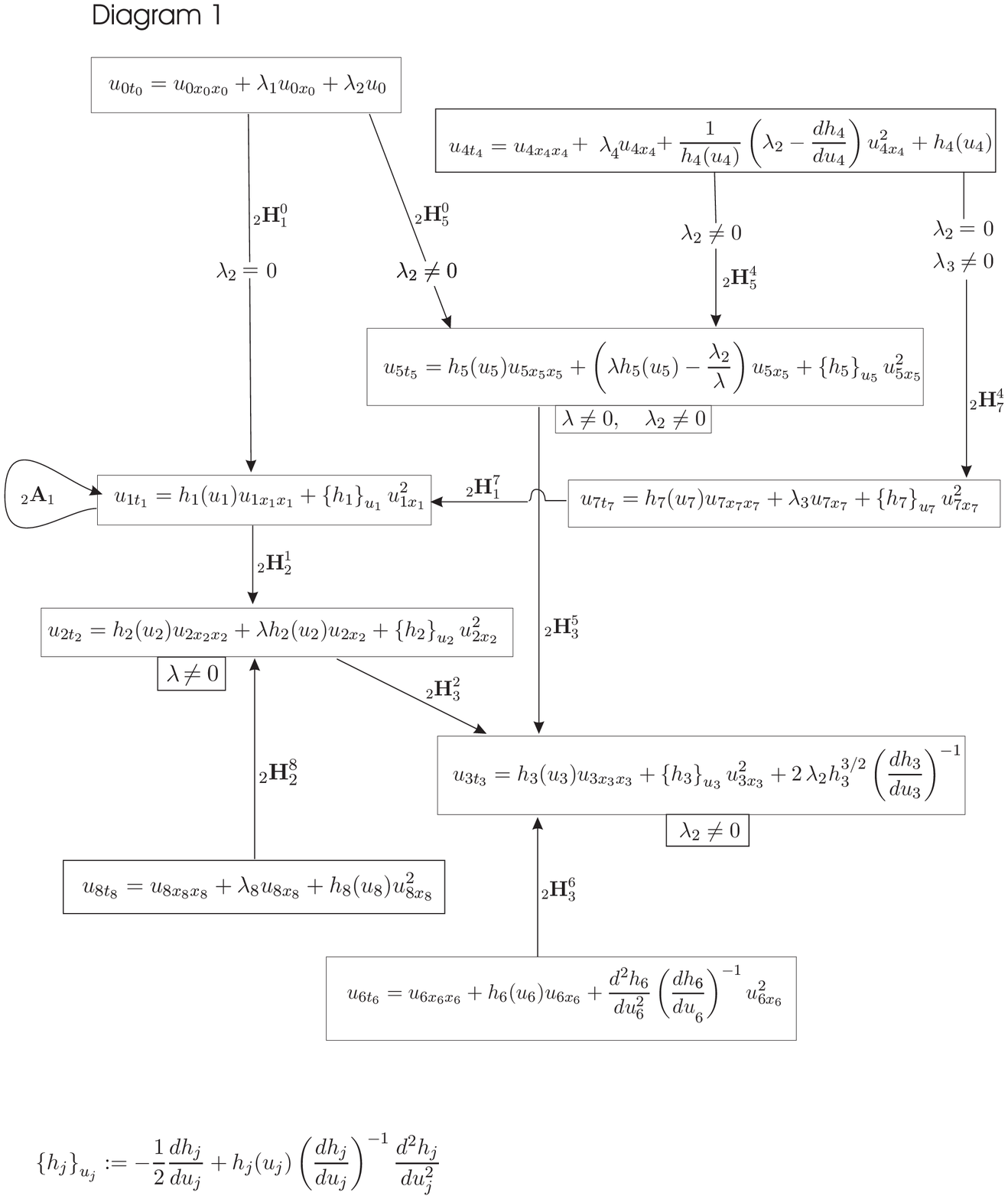}  
\end{center}
\end{figure}


The autohodograph transformation $_2\mbox{\bf A}_1$
which transforms (\ref{int_1}) into itself, i.e., in
\bg
\tilde u_{1\tTO}=h_1(\tilde u_1)\tilde u_{1\tXO \tXO}
+\left\{h_1\right\}_{\tilde u_1}{\tilde u_{1\tXO}}^2,
\ed
is given by
\bg
_2\mbox{\bf A}_1:\left\{\ba{l}
\displaystyle{dx_1(\tilde x_1,\tilde t_1)
=(\alpha \tilde x_1+\beta)
h_1^{-1/2}(\tilde u_1)d\tilde x_1
}\\[3mm]
\qquad\quad\displaystyle{+\left[\alpha h_1^{1/2}(\tilde u_1)
-\frac{1}{2}(\alpha \tilde x_1+\beta)
h_1^{-1/2}(\tilde u_1)\frac{dh_1}{d\tilde u_1}\tilde
u_{1\tilde x_1}\right]d\tilde t_1}
\\[3mm]
\displaystyle{dt_1(\tilde x_1,\tilde t_1)=d\tilde t_1}\\[3mm]
\displaystyle{h_1(u_1(x_1,t_1))=(\alpha \tilde x_1+\beta)^2,\ \alpha\in\Re\backslash\{0\},\ \beta\in\Re.
}
\ea\right.
\ed
It is noteworthy that (\ref{int_1}) is the only equation in Diagram 1
that admits an autohodograph transformation.

\strut\hfill

We consider three examples for the above Cases.

\strut\hfill

\noindent
{\bf Example 1:}
We consider Case III with $h_3=u_3^3$
and $\lambda_1=\lambda_2=1$, i.e.,
\beg
\label{ex_3}
u_{3\TTH}=u_3^3 u_{3\XTH\XTH}+\frac{1}{2}u_3^2u^2_{3\XTH}
+\frac{2}{3}u_3^{5/2}.
\eeq
It follows that (\ref{ex_3}) is linearised to
\beg
\label{ex_lin3}
u_{0\TN}=u_{0\XN\XN}+u_{0\XN}
\eeq
by the transformation
\begin{gather}
x_3(\XN,\TN)=\frac{2u_{0\XN\XN}}{u_{0\XN}}\notag\\
\label{ex_tr}
t_3(\XN,\TN)=t_0\\
u_3(\XTH,\TTH)=\left[\pde{\ }{\XN}\left(\frac{2u_{0\XN\XN}}{u_{0\XN}}
\right)\right]^{2/3}.\notag
\end{gather}
By group theoretical methods \cite{fush} we obtain the following solution
for (\ref{ex_lin3}):
\bg
u_0(\XN,\TN)=t_0^{-1/2}\exp\left[-\frac{1}{4}\left(\frac{\XN}{\TN}+1\right)^2
\TN\right].
\ed
Using (\ref{ex_tr}) we transform this solution into a solution for
(\ref{ex_3}), namely
\bg
u_3(\XTH,\TTH)=(-1)^{2/3}
\left[\frac{A +t_3)^2 +2\TTH}{A^2\TTH}
\right]^{2/3},
\ed
where
\bg
A= \frac{1}{2}\left(\XTH^2\TTH^2+8\TTH\right)^{1/2}
-\frac{1}{2}\XTH\TTH-\TTH.
\ed

\strut\hfill

\noindent
{\bf Example 2:} In Cases I, II, III we
let $\lambda_1=\lambda_2=0$, $\lambda=1$,
\bg
h_j(u_j)=u_j^2,\qquad j=1,2,3
\ed
and derive the transformation $_2\mbox{\bf H}_1^6$, which
transforms (\ref{int_6}) into
\beg
\label{ex_auto}
u_{1\TO}=u_1^2u_{1\XO\XO}.
\eeq
With the above assumptions,
(\ref{int_2}) and (\ref{int_3}) take the following respective forms:
\begin{gather}
\label{ex1_tw}
u_{2\TTW}=u_2^2u_{2\XTW\XTW}+u_2^2u_{2\XTW},\\[4mm]
u_{3\TTH}=u_3^2u_{3\XTH\XTH}+u_3^2.\notag
\end{gather}
The transformation into the autohodograph
invariant equation (\ref{ex_auto}) is obtained by the
composition
\bg
_2\mbox{\bf H}_1^6=\left(_2\mbox{\bf H}_3^1\right)^{-1}
\circ\,{_2}\mbox{\bf H}_3^6,
\ed
where
$\left(_2\mbox{\bf H}_3^1\right)^{-1}$ denotes the inverse transformation
of $_2\mbox{\bf H}_3^1$. In particular $_2\mbox{\bf H}_3^1
=\, {_2}\mbox{\bf H}_3^2\circ\,{_2}\mbox{\bf H}_2^1$. Then
\bg
_2\mbox{\bf H}_3^1:\left\{\ba{l}
\displaystyle{u_1^{-1}(x_1,t_1)dx_1(x_3,t_3)
= u_3^{-1}dx_3-\left(u_{3x_3}+\frac{1}{2}x_3\right)dt_3 }
\\[3mm]
dt_1(x_3,t_3)=dt_3\\[3mm]
\displaystyle{u_1(x_1,t_1) u_{1x_1x_1}(x_1,t_1)=\frac{1}{2}u_3 },
\ea\right.
\ed
so that
\bg
(_2\mbox{\bf H}_3^1)^{-1}:\left\{\ba{l}
\displaystyle{x_3(x_1,t_1)
=2u_{1x_1} }
\\[3mm]
dt_3(x_1,t_1)=dt_1\\[3mm]
\displaystyle{u_3(x_3,t_3)=2u_1 u_{1x_1x_1} }
\ea\right.
\ed
and
\bg
_2\mbox{\bf H}_1^6:\left\{\ba{l}
\displaystyle{dx_6(x_1,t_1)
=u_1^{-1}dx_1 -u_{1x_1}dt_1  }
\\[3mm]
dt_6(x_1,t_1)=dt_1\\[3mm]
\displaystyle{h_6(u_6(x_6,t_6))=2 u_{1x_1} }.
\ea\right.
\ed
The autohodograph transformation $_2\mbox{\bf A}_1$
transforming (\ref{ex_auto})
into
\bg
\td u_{1\tTO}=\td u_1^2\td u_{1\tXO\tXO}
\ed
(with $\alpha=1,\ \beta=0$)
follows, viz.
\bg
_2\mbox{\bf A}_1:\left\{\ba{l}
\displaystyle{dx_1(\td x_1,\td t_1)
=\td x_1\td u_1^{-1}d\td x_1 +\left(\td u_1
-\td x_1\td u_{1\td x_1}\right)d\td t_1  }
\\[3mm]
dt_1(\td x_1,\td t_1)=d\td t_1\\[3mm]
\displaystyle{u_1(x_1,t_1)=\td x_1 }.
\ea\right.
\ed
The inverse of $_2\mbox{\bf A}_1$ is
\bg
(_2\mbox{\bf A}_1)^{-1}:\left\{\ba{l}
\displaystyle{\td x_1=u_1}
\\[3mm]
d\td t_1=dt_1\\[3mm]
\displaystyle{\td u_1(\td x_1,\td t_1)=u_1u_{1x_1}. }
\ea\right.
\ed
The linearising transformation $_2\mbox{\bf H}_0^6$ of
(\ref{int_6}) into
\bg
u_{0\TN}=u_{0\XN\XN}
\ed
is obtained by
the composition
\bg
_2\mbox{\bf H}_0^6=\left(_2\mbox{\bf H}_1^0\right)^{-1}\circ
\left(_2\mbox{\bf H}_3^1\right)^{-1}\circ\,{_2}\mbox{\bf H}_3^6\ ,
\ed
where
\bg
\left(_2\mbox{\bf H}_1^0\right)^{-1}:\left\{\ba{l}
\displaystyle{x_1(x_0,t_0)=u_0}
\\[3mm]
dt_1(x_0,t_0)=dt_0\\[3mm]
\displaystyle{u_1(x_1,t_1)=u_{0x_0} }
\ea\right.
\ed
and
\bg
_2\mbox{\bf H}_3^6:\left\{\ba{l}
\displaystyle{dx_6(x_3,t_3)=u_3^{-1}dx_3-(u_{3x_3}+x_3)dt_3}
\\[3mm]
dt_6(x_3,t_3)=dt_3\\[3mm]
\displaystyle{h_6(u_6(x_6,t_6))=x_3 }.
\ea\right.
\ed
Note that, with $h_6(u_6)=u_6$ (\ref{int_6}) reduces to the Burgers'
equation and $_2\mbox{\bf H}_0^6$ becomes the Cole-Hopf transformation, i.e.,
\bg
_2\mbox{\bf H}_0^6:\left\{\ba{l}
\displaystyle{x_6(x_0,t_0)=x_0}
\\[3mm]
t_6(x_0,t_0)=t_0\\[3mm]
\displaystyle{u_6(x_6,t_6)=2\phi^{-1}(x_0,t_0)\pde{\phi}{x_0}(x_0,t_0) },
\ea\right.
\ed
where $\phi(x_0,t_0)=\p u_0/\p x_0$.


\strut\hfill

\noindent
{\bf Example 3:} Consider Cases III, V and
let $\lambda=\lambda_1=\lambda_2=1$
\bg
h_j(u_j)=u_j^2,\qquad j=3,5.
\ed
We derive the transformation $_2\mbox{\bf H}_0^6$ which
linearises (\ref{int_6}), i.e.,
\bg
u_{6\TS}=u_{6\XS\XS}+h_6(u_6)u_{6\XS}
+\frac{d^2h_6}{du_6^2}\left(\frac{dh_6}{du_6}\right)^{-1}u^2_{6\XS}
\ed
into
\bg
u_{0\TN}=u_{0\XN\XN}+u_{0\XN}+u_0.
\ed
This linearisation is obtained by the following compositions:
\bg
_2\mbox{\bf H}_0^6=(_2\mbox{\bf H}_3^0)^{-1}
\circ\, {_2}\mbox{\bf H}_3^6\quad\mbox{with}\quad
_2\mbox{\bf H}_3^0=\, {_2}\mbox{\bf H}_3^5\circ\,{_2}\mbox{\bf H}_5^0.
\ed
Under the above assumptions (\ref{int_5}) takes the form
\beg
\label{ex2_f}
u_{5\TV}=u_5^2 u_{5\XV\XV}
+\left(u_5^2-1\right)u_{5\XV}
\eeq
and (\ref{int_3}) is 
\beg
\label{ex2_new}
u_{3\TTH}=u_3^2u_{3\XTH\XTH}+u_3^2.
\eeq
The linearising transformation for (\ref{int_6}) is then
\bg
_2\mbox{\bf H}_0^6:\left\{\ba{l}
\displaystyle{dx_6(x_0,t_0)=dx_0+dt_0}
\\[3mm]
dt_6(x_0,t_0)=dt_0\\[3mm]
\displaystyle{h_6(u_6(x_6,t_6))=2u_0^{-1} u_{0x_0} },
\ea\right.
\ed
whereas (\ref{ex2_f}) is linearised by 
\bg
(_2\mbox{\bf H}_5^0)^{-1}:\left\{\ba{l}
\displaystyle{x_5(x_0,t_0)=\ln|u_0|}
\\[3mm]
dt_5(x_0,t_0)=dt_0\\[3mm]
\displaystyle{u_5(x_5,t_5)=u_0^{-1}u_{0x_0} }
\ea\right.
\ed
and (\ref{ex2_new}) by
\bg
(_2\mbox{\bf H}_3^0)^{-1}:\left\{\ba{l}
\displaystyle{x_3(x_0,t_0)=2u_0^{-1}u_{0x_0}   }
\\[3mm]
dt_3(x_0,t_0)=dt_0\\[3mm]
\displaystyle{u_3(x_3,t_3)=2u_0^{-1}u_{0x_0x_0}-2u_0^{-2}
u^2_{0x_0}. }
\ea\right.
\ed

\noindent
{\bf Remark:} The linearisation of (\ref{ex2_new})
is also given in \cite{sok}.

\section{Linearisable nonautonomous second-order equations}

Next we list the nonautonomous second-order evolution equations which
have been constructed using (\ref{2nd_hodo}), as well as their
corresponding linearising transformations. Each equation
given in Cases I -- VIII above results, by (\ref{2nd_hodo}),
in a nonautonomous equation leading to eight further cases.
Diagram 2 shows the connection to the autonomous cases.

\strut\hfill

\noindent
{\bf Case IX:} Let $\lambda_1\in \Re$, 
$\{h_1,k_1\}\in C^2(\Re)$ with $dh_1/d\tXO\neq 0$ and
$dk_1/d\tilde u_1\neq 0$.
Then
\begin{gather}
\label{int_9}
\tilde u_{1\tTO}=k_1(\tilde u_1)\tilde u_{1\tXO\tXO}
+\frac{k_1(\tilde u_1)\left\{h_1\right\}_{\tXO}}
{h_1(\tXO)}\tilde u_{1\tXO}
+\left\{k_1\right\}_{\tone}\tilde u^2_{1\tXO}\notag\\[3mm]
\qquad
+2k_1^2(\tilde u_1)\left(\frac{dk_1}{d\tilde u_1}\right)^{-1}\frac{d}{d\tXO}
\left(\frac{\left\{h_1\right\}_{\tXO}}{h_1(\tXO)}\right)
\end{gather}
is linearised to $u_{0\TN}=u_{0\XN\XN}+\lambda_1 u_{0\XN}$ by the
transformation
\bg
_2{\mbox{\bf L}}_0^{\tilde 1}:\left\{\ba{l}
\displaystyle{h_1(\tilde x_1(x_0,t_0))=u_{0\XN}^2}
\\[3mm]
d\tilde t_1(x_0,t_0)=dt_0\\[3mm]
\displaystyle{k_1(\tilde u_1(\tXO,\tTO))=\left.4u^2_{0\XN}u^2_{0\XN\XN}
\left[
\left(
\frac{dh_1}{d\tXO}\right)^{-2}\right]\right|_{h_1(\tilde
x_1)=u^2_{0x_0}}.
}
\ea\right.
\ed

\vspace{4mm}

\noindent
{\bf Case X:} Let $\lambda\in\Re\backslash \{0\}$,
$\lambda_1\in \Re$,
$\{h_2,k_2\}\in C^2(\Re)$ with $dh_2/d\tXTW\neq 0$ and
$dk_2/d\tilde u_2\neq 0$.
Then
\begin{gather}
\label{int_10}
\tilde u_{2\tTTW}=k_2(\tilde u_2)\tilde u_{2\tXTW\tXTW}
+\left\{k_2\right\}_{\ttwo}\tilde u^2_{2\tXTW}
+2k_2^2(\tilde u_2)\left(\frac{dk_2}{d\tilde u_2}\right)^{-1}
\frac{d}{d\tXTW}\left(\frac{\left\{h_2\right\}_{\tXTW}}{h_2(\tXTW)}\right)
\notag\\[3mm]
\qquad
+\frac{k_2(\tilde u_2)\left\{h_2\right\}_{\tXTW}}{h_2(\tXTW)}
\tilde u_{2\tXTW}
+2\lambda h_2^{-1/2}(\tilde x_2)\frac{dh_2}{d\tXTW}
\left(\frac{dk_2}{d\tilde u_2}\right)^{-1} k_2^{3/2}(\tilde u_2)
\end{gather}
is linearised to $u_{0\TN}=u_{0\XN\XN}+\lambda_1 u_{0\XN}$ by the
transformation
\bg
_2{\mbox{\bf L}}_0^{\tilde 2}:\left\{\ba{l}
\displaystyle{h_2(\tilde x_2(x_0,t_0))=\frac{1}{\lambda^2}
\left(\frac{u_{0\XN\XN}}{u_{0\XN}}\right)^2
}
\\[3mm]
d\tilde t_2(x_0,t_0)=dt_0\\[3mm]
\displaystyle{k_2(\tilde
u_2(\tXTW,\tTTW))=\frac{4}{\lambda^4}
\left(\frac{u_{0\XN\XN}}{u_{0\XN}}\right)^2
\left[\frac{\p}{\p \XN}
\left(\frac{u_{0\XN\XN}}{u_{0\XN}}\right)\right]^2\times
}\\[4mm]
\qquad\qquad\qquad\qquad\qquad\qquad\qquad\qquad\displaystyle{
\left.
\times\left[
\left(\frac{dh_2}{d\tXTW}\right)^{-2}
\right]
\right|_{
h_2(\tilde x_2)=\frac{1}{\lambda^2}
\left(\frac{u_{0x_0x_0}}{u_{0x_0}}\right)^2 }.}
\ea\right.
\ed

\vspace{4mm}

\noindent
{\bf Case XI:} Let $\lambda_1\in\Re$, $\lambda_2\in\Re\backslash
\{0\}$, 
$\{h_3,k_3\}\in C^2(\Re)$ with $dh_3/d\tXTH\neq 0$ and
$dk_3/d\tilde u_3\neq 0$.
Then 
\begin{gather}
\label{int_11}
 \tilde u_{3\tTTH}=k_3(\tilde u_3)\tilde u_{3\tXTH\tXTH}
+\left(\frac{k_3(\tilde u_3)\left\{h_3\right\}_{\tXTH} }{h_3(\tXTH)}
-2\lambda_2 h_3^{3/2}(\tXTH)
\left(\frac{dh_3}{d \tXTH}\right)^{-1}\right) \tilde u_{3\tXTH}\notag\\[2mm]
\qquad
+\left\{k_3\right\}_{\tthree} \tilde u^2_{3\tXTH}\notag
+2\lambda_2k_3(\tilde u_3)\left(\frac{dk_3}{d\tilde u_3}\right)^{-1}
\left(4h_3^{1/2}(\tXTH)-2h_3^{3/2}(\tXTH)
\left(\frac{dh_3}{d\tXTH}\right)^{-2}
\frac{d^2 h_3}{d\tXTH^2}\right)\notag\\[2mm]
\qquad
+2k_3^2(\tilde u_3)\left(\frac{dk_3}{d\tilde u_3}\right)^{-1}
\frac{d}{d \tXTH}
\left(\frac{\left\{h_3\right\}_{\tXTH}}{h_3(\tXTH)}
\right)
\end{gather}
is linearised to
$u_{0\TN}=u_{0\XN\XN}+\lambda_1 u_{0\XN}$ by the
transformation
\bg
_2{\mbox{\bf L}}_0^{\tilde 3}:\left\{\ba{l}
\displaystyle{h_3(\tilde x_3(x_0,t_0))=\frac{4}{\lambda_2^2}
\left[\frac{\p}{\p \XN}\left(\frac{u_{0\XN\XN}}{u_{0\XN}}\right)\right]^2
}
\\[3mm]
d\tilde t_3(x_0,t_0)=dt_0\\[3mm]
\displaystyle{k_3(\tilde u_3(\tXTH,\tTTH))=
\frac{64}{\lambda_2^4}
\left[\pde{\ }{\XN}\left(\frac{u_{0\XN\XN}}{u_{0\XN}}\right)\right]^2
\left[\pdd{\ }{\XN}\left(\frac{u_{0\XN\XN}}{u_{0\XN}}\right)\right]^2
\times
}
\\[5mm]
\qquad\qquad\qquad\qquad\qquad\qquad\qquad\displaystyle{
\left.
\times\left[
\left(\frac{dh_3}{d\tXTH}\right)^{-2}
\right]
\right|_{
h_3(\tilde x_3)=\frac{4}{\lambda_2^2}
\left[\frac{\p}{\p x_0}\left(\frac{u_{0x_0x_0}}{u_{0x_0}}\right)\right]^2 }.}
\ea\right.
\ed

\vspace{4mm}

\noindent
{\bf Case XII.1:} Let $\lambda_1\in\Re$,
$\lambda_2\in \Re\backslash\{0\}$, 
$\{h_4,k_4\}\in C^2(\Re)$ with $dh_4/d\tXF\neq 0$ and
$dk_4/d\tilde u_4\neq 0$.
Then 
\begin{gather}
\label{int_12}
\tilde u_{4\tTF}=k_4(\tilde u_4)\tilde u_{4\tXF\tXF}
+\left[\vphantom{\frac{dk_4}{du_4}}
\left(\frac{\lambda_2}{h_4(\tXF)}-\frac{1}{h_4(\tXF)}\frac{dh_4}{d\tXF}
\right)k_4(\tilde u_4)
-h_4(\tXF)\right]\tilde u_{4\tXF}\notag\\[3mm]
\qquad
+\left\{k_4\right\}_{\tfour} \tilde u^2_{4\tXF}
+2\left[k_4^2(\tilde u_4)
\left(-\frac{\lambda_2}{h_4^2(\tXF)}\frac{dh_4}{d\tXF}-\frac{1}{h_4(\tXF)}
\frac{d^2h_4}{d\tXF^2}+\frac{1}{h_4^2(\tXF)}
\left(\frac{dh_4}{d\tXF}\right)^2\right)\right.\notag\\[3mm]
\qquad\left.\vphantom{\left(\left(\frac{dh_4}{d\tXF}\right)^2   \right)}
+k_4(\tilde u_4)
\frac{dh_4}{d\tXF}\right]
\left(\frac{dk_4}{d\tilde u_4}\right)^{-1}
\end{gather}
is linearised to $u_{0\TN}=u_{0\XN\XN}+\lambda_1 u_{0\XN}+\lambda_2 u_0$ by the
transformation
\bg
_2{\mbox{\bf L}}_0^{\tilde 4.1}:\left\{\ba{l}
\displaystyle{\int^{\tilde x_4(x_0,t_0)}\frac{d\xi}{h_4(\xi)}
=\frac{1}{\lambda_2}\ln | u_0|
}
\\[5mm]
d\tilde t_4(x_0,t_0)=dt_0\\[3mm]
\displaystyle{k_4(\tilde
u_4(\tXF,\tTF))=\left.\frac{1}{\lambda_2^2}
\left(\frac{u_{0\XN}}{u_0}\right)^2
\left[\vphantom{\left(\frac{u_{0\XN}}{u_0}\right)^2}
h_4^2(\tXF)\right]
\right|_{\int^{\tilde x_4}\frac{d\xi}{h_4(\xi)}
=\frac{1}{\lambda_2}\ln|u_0|}.
}
\ea\right.
\ed

\noindent
{\bf Case XII.2:} Let $\lambda_1\in\Re$,
$\lambda_3\in\Re\backslash\{0\}$,
$\{h_4,k_4\}\in C^2(\Re)$ with $dh_4/d\tXF\neq 0$ and
$dk_4/d\tilde u_4\neq 0$.
Then 
\begin{gather}
\label{int_12.2}
\tilde u_{4\tTF}=k_4(\tilde u_4)\tilde u_{4\tXF\tXF}
-\left(\vphantom{\frac{dk_4}{du_4}}
\frac{k_4(\tilde u_4)}{h_4(\tXF)}\frac{dh_4}{d\tXF}
+h_4(\tXF)\right)\tilde u_{4\tXF}+\left\{k_4\right\}_{\tfour} \tilde
u^2_{4\tXF}
\notag\\[3mm]
\qquad
+2\left[k_4^2(\tilde u_4)
\left(-\frac{1}{h_4(\tXF)}
\frac{d^2h_4}{d\tXF^2}+\frac{1}{h_4^2(\tXF)}
\left(\frac{dh_4}{d\tXF}\right)^2\right)
+k_4(\tilde u_4)
\frac{dh_4}{d\tXF}\right]
\left(\frac{dk_4}{d\tilde u_4}\right)^{-1}
\end{gather}
is linearised to $u_{0\TN}=u_{0\XN\XN}+\lambda_1 u_{0\XN}$ by the
transformation
\bg
_2{\mbox{\bf L}}_0^{\tilde 4.2}:\left\{\ba{l}
\displaystyle{

h_4^{-1}(\tilde x_4)d\tilde x_4(\XN,\TN)=-\frac{1}{\lambda_3}u_0d\XN
-\frac{1}{\lambda_3}\left(u_{0\XN}+\lambda_1u_0-\lambda_3\right)d\TN
--- (*)
}
\\[5mm]
d\tilde t_4(x_0,t_0)=dt_0\\[3mm]
\displaystyle{k_4(\tilde
u_4(\tXF,\tTF))=\left.\frac{u_0^2}{\lambda_3^2}
\left[\vphantom{\frac{1}{\lambda}}
h_4^2(\tilde x_4)\right]\right|_{(*)}.
}
\ea\right.
\ed
Remark: By $\displaystyle \left.\vphantom{\frac{df}{dx}}\right|_{(*)}$
we mean that the function $h_4(\tilde x_4)$ has to be written
in terms of $x_0$, $t_0$, $u_0$ and its derivatives with respect
to $x_0$, obtained from the expression $(*)$.

\vspace{4mm}

\noindent
{\bf Case XIII:} Let $\lambda_1\in\Re$,
$\{\lambda,\lambda_2\}\in\Re\backslash\{0\}$,
$\{h_5,k_5\}\in C^2(\Re)$ with $dh_5/d\tXV\neq 0$ and
$dk_5/d\tilde u_5\neq 0$.
Then 
\begin{gather}
\label{int_13}
\tilde u_{5\tTV}=k_5(\tilde u_5)\tilde u_{5\tXV\tXV}
+\frac{k_5(\tilde u_5)\{h_5\}_{\tXV}}{h_5(\tXV)}\tilde u_{5\tXV}
+\{k_5\}_{\five}\tilde u^2_{5\tXV}\notag\\[3mm]
\qquad
+2k_5^2(\tilde u_5)
\left(\frac{dk_5}{d\tilde u_5}\right)^{-1}
\frac{d}{d\tXV}\left(\frac{\{h_5\}_{\tXV}}{h_5(\tXV)}\right)
+2\lambda h_5^{-1/2}(\tXV)
\frac{dh_5}{d\tXV}
k_5^{3/2}(\tilde u_5)
\left(\frac{dk_5}{d\tilde u_5}\right)^{-1}
\end{gather}
is linearised to $u_{0\TN}=u_{0\XN\XN}+\lambda_1 u_{0\XN}+\lambda_2 u_0$ by the
transformation
\bg
_2{\mbox{\bf L}}_0^{\tilde 5}:\left\{\ba{l}
\displaystyle{h_5(\tilde x_5(x_0,t_0))=\frac{1}{\lambda^2}
\left(\frac{u_{0\XN}}{u_0}\right)^2
}
\\[3mm]
d\tilde t_5(x_0,t_0)=dt_0\\[3mm]
\displaystyle{k_5(\tilde u_5(\tXV,\tTV))=\frac{1}{\lambda^2}
\left[\frac{\p }{\p \XN}\left(\frac{u_{0\XN}}{u_0}\right)^2\right]^2
\left.
\left[
\left(\frac{dh_5}{d\tXV}\right)^{-2}
\right]
\right|_{
h_5(\tilde x_5)=\frac{1}{\lambda^2}
\left(\frac{u_{0x_0}}{u_{0}}\right)^2 }.
}
\ea\right.
\ed

\vspace{4mm}

\noindent
{\bf Case XIV:} Let $\lambda_1\in\Re$, $\lambda_2\in\Re\backslash\{0\}$,
and $\{h_6,k_6\}\in C^2(\Re)$ with $dh_6/d\tXS\neq 0$ and
$dk_6/d\tilde u_6\neq 0$.
Then 
\begin{gather}
\label{int_14}
\tilde u_{6\tTS}=k_6(\tilde u_6)\tilde u_{6\tXS\tXS}
+k_6(\tilde u_6)\frac{d^2 h_6}{d\tXS^2}
\left(\frac{dh_6}{d\tXS}\right)^{-1} \tilde u_{6\tXS}
+\{h_6\}_{\tsix} \tilde u^2_{\tXS}\notag\\[3mm]
\qquad
+2k_6^2(\tilde u_6)\left(\frac{dk_6}{d\tilde u_6}\right)^{-1}
\frac{d}{d\tXS}\left[\frac{d^2 h_6}{d\tXS^2}
\left(\frac{dh_6}{d\tXS}\right)^{-1}\right]
+2k_6^{3/2}(\tilde u_6)\left(\frac{dk_6}{d\tilde u_6}\right)^{-1}
\frac{dh_6}{d\tXS}
\end{gather}
is linearised to $u_{0\TN}=u_{0\XN\XN}+\lambda_1 u_{0\XN}$ by the
transformation
\bg
_2{\mbox{\bf L}}_0^{\tilde 6}:\left\{\ba{l}
\displaystyle{h_6(\tilde x_6(x_0,t_0))=
2\frac{u_{0\XN\XN}}{u_{0\XN}}
}
\\[3mm]
d\tilde t_6(x_0,t_0)=dt_0\\[3mm]
\displaystyle{k_6(\tilde u_6(\tXS,\tTS))=\left.4\left[\frac{\p}{\p
\XN}\left(\frac{u_{0\XN\XN}}{u_{0\XN}}\right)\right]^2
\left[
\left(\frac{dh_6}{d\tXS}\right)^{-2}\right]
\right|_{h_6(\tilde x_6)=
2\frac{u_{0x_0x_0}}{u_{0x_0}}}.
}
\ea\right.
\ed

\vspace{4mm}

\noindent
{\bf Case XV:} Let $\{\lambda_1,\lambda_3\}\in\Re$,
$\{h_7,k_7\}\in C^2(\Re)$ with $dh_7/d\tXSV\neq 0$ and
$dk_7/d\tilde u_7\neq 0$.
Then 
\begin{gather}
\label{int_15}
\tilde u_{7\tTSV}=k_7(\tilde u_7)\tilde u_{7\tXSV\tXSV}
+\frac{k_7(\tilde u_7)\{h_7\}_{\tXSV}}{h_7(\tXSV)} \tilde u_{7\tXSV}
+\{k_7\}_{\tseven}\tilde u_{7\tXSV}^2\notag\\[3mm]
\qquad +2k_7^2(\tilde u_7)\left(\frac{dk_7}{d\tilde u_7}\right)^{-1}
\frac{d}{d\tXSV}\left(\frac{\{h_7\}_{\tXSV}}{h_7(\tXSV)}\right)
\end{gather}
is linearised to $u_{0\TN}=u_{0\XN\XN}+\lambda_1 u_{0\XN}$ by the
transformation
\bg
_2{\mbox{\bf L}}_0^{\tilde 7}:\left\{\ba{l}
\displaystyle{h_7(\tilde x_7(x_0,t_0))=u_0^2
}
\\[3mm]
d\tilde t_7(x_0,t_0)=dt_0\\[3mm]
\displaystyle{k_7(\tilde u_7(\tXSV,\tTSV))=
\left. 4u_{0}^2u_{0\XN}^2
\left[\left(\frac{dh_7}{d\tXSV}\right)^{-2}\right]
\right|_{h_7(\tilde x_7)=u_0^2}.
}
\ea\right.
\ed

\vspace{4mm}

\noindent
{\bf Case XVI:} Let $\lambda\in\Re\backslash\{0\}$,
$\lambda_1\in\Re$,
$\{h_8,k_8\}\in C^2(\Re)$ with
$dk_8/d\tilde u_8\neq 0$.
Then 
\begin{gather}
\label{int_16}
\tilde u_{8\tTE}=k_8(\tilde u_8)\tilde u_{8\tXE\tXE}
+k_8(\tilde u_8)h_8(\tXE)\tilde u_{8\tXE}+\{k_8\}_{\teight}\tilde
u_{8\tXE}^2\notag\\[3mm]
\qquad
+2k_8^2(\tilde u_8)\frac{dh_8}{d\tXE}\left(\frac{dk_8}{d\tilde
u_8}\right)^{-1}
\end{gather}
is linearised to $u_{0\TN}=u_{0\XN\XN}+\lambda_1 u_{0\XN}$ by the
transformation
\bg
_2{\mbox{\bf L}}_0^{\tilde 8}:\left\{\ba{l}
\ba[b]{c}
\scriptstyle \tXE\\
\bigint
\ea
\displaystyle{
\left\{\exp\left(\int^\xi
h_8(\eta)d\eta\right) \left[\lambda\int^\xi\exp\left(\int^{\eta}
h_8(\eta')d\eta'\right)d\eta\right]^{-1}\right\}d\xi
}\\[8mm]
\displaystyle{\hphantom{d\tilde t_8(x_0,t_0)}
=\frac{1}{2\lambda}\ln\left|u_{0\XN}^2\right| ---(*)
}
\\[5mm]
d\tilde t_8(x_0,t_0)=dt_0\\[3mm]
\displaystyle{k_8(\tilde u_8(\tXE,\tTE))=\left.\frac{u^2_{0\XN\XN}}{u^2_{0\XN}}
\left[\frac{\int^{\tXE}\left[\exp\left(\int^{\xi}
h_8(\eta)d\eta\right)\right]d\xi}{\exp\left(\int^{\tXE}h_8(\xi)d\xi\right)}
\right]^2\right|_{(*)}.
}
\ea\right.
\ed


\begin{figure}
\begin{center}
\includegraphics[width=15cm]{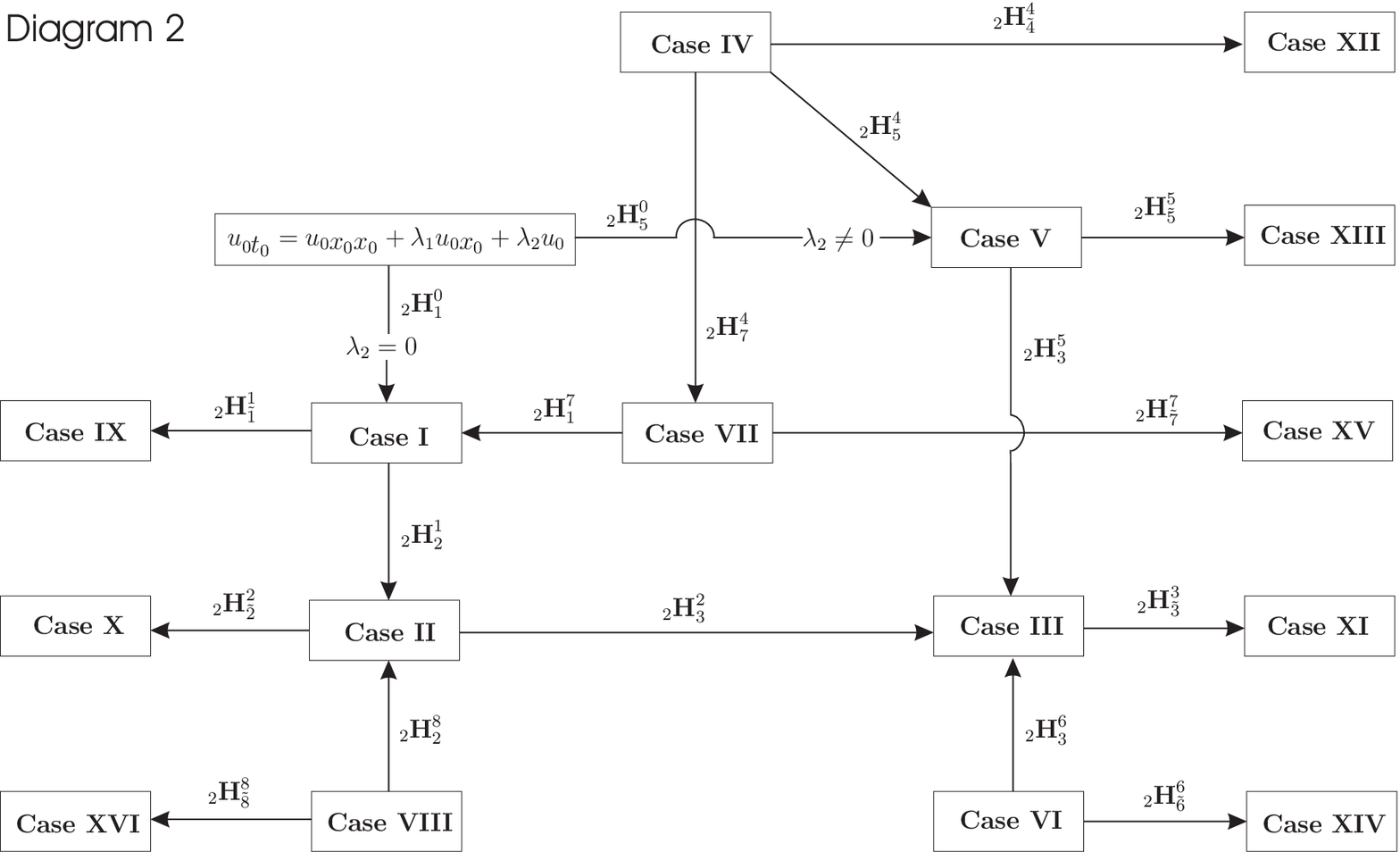}  
\end{center}
\end{figure}

\strut\hfill

\noindent
{\bf Example 4:} We consider Case XI with
$h_3(\tXTH) =\tXTH^2$, $k_3(\tilde u_3)=\tilde u_3^2$
and $\lambda_2=-1$, i.e.,
\beg
\label{ex_4}
\tilde u_{3\tTTH}=\tilde u^2_3 \tilde u_{3\tXTH\tXTH}
+\tXTH^2 \tilde u_{3\tXTH}-3\tXTH \tilde u_3.
\eeq
It follows that (\ref{ex_4}) is linearised to
\bg
\label{ex_lin4}
u_{0\TN}=u_{0\XN\XN}+\lambda_1 u_{0\XN}
\ed
by the transformation
\begin{gather}
\label{ex_trans4}
\tilde x_3(\XN,\TN)=\pde{\ }{\XN}\left(
\frac{-2u_{0\XN\XN}}{u_{0\XN}}\right)
\notag\\
 d\tilde t_3(\XN,\TN)=dt_0\notag\\
\tilde u_3(\tXTH,\tTTH)=\pdd{\ }{\XN}\left(\frac{-2u_{0\XN\XN}}{u_{0\XN}}
\right).\notag
\end{gather}
{\bf Remark:} Equation (\ref{ex_4}) was proposed
in \cite{sok}.

\strut\hfill

\noindent
{\bf Example 5:} We consider Case XVI with $h_8(\tXE)=\tXE^{-1}$
and $k_8(\tilde u_8)=\tilde u_8$, i.e.,
\beg
\label{ex_5}
\tilde u_{8\tXE}=\tilde u_8\tilde u_{8\tXE\tXE}
+\tXE^{-1}\tilde u_8\tilde u_{8\tXE}
-\frac{1}{2}\tilde u_{8\tXE}^2-2\tXE^{-2}\tilde u_8^2
\eeq
It follows that (\ref{ex_5}) is linearised to
\bg
\label{ex_lin5}
u_{0\TN}=u_{0\XN\XN}+\lambda_1 u_{0\XN}
\ed
by the transformation
\begin{gather}
\tXE(\XN,\TN)=u_{0\XN}^{1/2}\notag\\
d\tTE=d\TN\notag\\
\tilde u_8(\tXE,\tTE)=\frac{1}{4}\frac{u^2_{0\XN\XN}}{u_{0\XN}}.\notag
\end{gather}

\strut\hfill

\noindent
{\large {\bf Acknowledgment:}}
We thank R Conte and P G L Leach for their comments on
the preprint version of this paper. M.E was 
financially supported by a Wallenberg Research Grant.
We also thank Johan Bystr\"om for helping us typeset
the Diagrams.




\strut\hfill

\strut\hfill

\noindent
{\Large{\bf Appendix}}

\strut\hfill

\noindent
Below we list the $x$-generalised
hodograph transformations $_2\mbox{\bf H}_j^i$
corresponding to Cases I -- VIII.
Diagram 1 shows the direction in which the transformations
act.

\bg
_2\mbox{\bf H}_1^0:\left\{\ba{l}
\displaystyle{dx_0(x_1,t_1)=h_1^{-1/2}(u_1)dx_1+\left(-\frac{1}{2}h_1^{-1/2}(u_1)
\frac{dh_1}{du_1}u_{1x_1}-\lambda_1\right)dt_1}
\\[3mm]
dt_0(x_1,t_1)=dt_1\\[3mm]
u_0(x_0,t_0)=\alpha_1 x_1+\beta_1,\qquad \alpha_1\in\Re\backslash\{0\},\
\beta_1\in\Re
\ea\right.
\ed

\bg
_2\mbox{\bf H}_5^0:\left\{\ba{l}
\displaystyle{dx_0(x_5,t_5)
=h_5^{-1/2}(u_5)dx_5+\left[-\lambda h_5^{1/2}(u_5)
\phantom{\frac{\lambda_1}{\lambda} 1234567890123456789}\right.}\\[3mm]
\qquad\quad\displaystyle{\left.-\frac{\lambda_2}{\lambda}h_5^{-1/2}(u_5)
-\frac{1}{2}h_5^{-1/2}(u_5)\frac{dh_5}{du_5}u_{5x_5}-\lambda_1
\right]dt_5}
\\[3mm]
dt_0(x_5,t_5)=dt_5\\[3mm]
\displaystyle{u_0(x_0,t_0)=\frac{1}{\lambda}e^{\lambda x_5+\alpha_5}},
\qquad \lambda\in\Re\backslash\{0\},\ \alpha_5\in\Re
\ea\right.\hfill
\ed

\bg
_2\mbox{\bf H}_2^1:\left\{\ba{l}
\displaystyle{dx_1(x_2,t_2)
=e^{\lambda x_2}\sqrt{\beta_2}h_2^{-1/2}(u_2)dx_2
-\frac{1}{2}e^{\lambda x_2}\sqrt{\beta_2}h_2^{-1/2}(u_2)
\frac{dh_2}{du_2}u_{2x_2}dt_2}
\\[3mm]
dt_1(x_2,t_2)=dt_2\\[3mm]
h_1\left(u_1(x_1,t_1)\right)=\beta_2 e^{2\lambda x_2},
\qquad \lambda\in\Re\backslash\{0\},\ \beta_2\in\Re\backslash\{0\}
\ea\right.
\ed



\bg
_2\mbox{\bf H}_3^2:\left\{\ba{l}
\displaystyle{dx_2(x_3,t_3)
=\left(\frac{\lambda_2}{2\lambda} x_3+\beta_3\right)h_3^{-1/2}(u_3)dx_3}\\[3mm]
\qquad\quad\displaystyle{+
\left[\frac{\lambda_2}{2\lambda}h_3^{1/2}(u_3)
-\frac{1}{2}(\frac{\lambda_2}{2\lambda}x_3+\beta_3)h_3^{-1/2}(u_3)
\frac{dh_3}{du_3}u_{3x_3}\right. }\\[3mm]
\displaystyle{\left.
\qquad\quad -\lambda(
\frac{\lambda_2}{2\lambda}x_3+\beta_3)^2\right]dt_3}
\\[3mm]
dt_2(x_3,t_3)=dt_3\\[3mm]
\displaystyle{h_2\left(u_2(x_2,t_2)\right)=\left(\frac{\lambda_2}{2\lambda}x_3
+\beta_3\right)^2,}
\qquad \{\lambda,\lambda_2\}\in\Re\backslash\{0\},\ 
\beta_3\in\Re
\ea\right.
\ed

\bg
_2\mbox{\bf H}_5^4:\left\{\ba{l}
\displaystyle{dx_4(x_5,t_5)
=h_5^{-1/2}(u_5)dx_5+\left[-\lambda h_5^{1/2}(u_5)
\phantom{\frac{\lambda_1}{\lambda}1234567890123456789}\right.}\\[3mm]
\qquad\quad\displaystyle{\left.-\frac{1}{2}h_5^{-1/2}(u_5)
\frac{dh_5}{du_5}u_{5x_5}-\frac{\lambda_2}{\lambda}h_5^{-1/2}(u_5)
-\lambda_4
\right]dt_5}
\\[3mm]
dt_4(x_5,t_5)=dt_5\\[3mm]
\displaystyle{
\int^{\displaystyle{u_4(x_4,t_4)}}\frac{d\xi}{h_4(\xi)}
=\frac{\lambda}{\lambda_2}\,x_5+\beta_5},
\quad \{\lambda,\lambda_2\}\in\Re\backslash\{0\},\
\{\beta_5,\lambda_4\}\in\Re
\ea\right.
\ed

\bg
_2\mbox{\bf H}_7^4:\left\{\ba{l}
\displaystyle{dx_4(x_7,t_7)
=h_7^{-1/2}(u_7)dx_7+\left[-\frac{1}{2}h_7^{-1/2}(u_7)
\frac{dh_7}{du_7}u_{7x_7}+\lambda_3h_7^{-1/2}(u_7)-\lambda_4
\right]dt_7}
\\[3mm]
dt_4(x_7,t_7)=dt_7\\[3mm]
\displaystyle{
\int^{\displaystyle{u_4(x_4,t_4)}}\frac{d\xi}{h_4(\xi)}
=-\frac{1}{\lambda_3}\,x_7+\beta_7},
\qquad \lambda_3\in\Re\backslash\{0\},\ \{\lambda_4,\beta_7\}\in\Re
\ea\right.
\ed

\bg
_2\mbox{\bf H}_3^5:\left\{\ba{l}
\displaystyle{dx_5(x_3,t_3)
=\left(\frac{\lambda_2}{2\lambda}x_3+\tilde \beta_3\right)
h_3^{-1/2}(u_3)dx_3+\left[\frac{\lambda_2}{2\lambda}
 h_3^{1/2}(u_3)
\phantom{\frac{\lambda_1}{\lambda}}\right.}\\[3mm]
\qquad\quad\displaystyle{\left.-\frac{1}{2}
\left(\frac{\lambda_2}{2\lambda}x_3+\tilde\beta_3\right)
h_3^{-1/2}(u_3)
\frac{dh_3}{du_3}u_{3x_3}
-\lambda\left(\frac{\lambda_2}{2\lambda}x_3+\tilde\beta_3\right)^2
+\frac{\lambda_2}{\lambda}
\right]dt_3}
\\[3mm]
dt_5(x_3,t_3)=dt_3\\[3mm]
\displaystyle{h_5\left(u_5(x_5,t_5)\right)=
\left(\frac{\lambda_2}{2\lambda}\,x_3+\tilde\beta_3\right)^2},
\qquad \{\lambda,\lambda_2\}\in\Re\backslash\{0\},\ 
\tilde\beta_3\in\Re
\ea\right.
\ed

\bg
_2\mbox{\bf H}_1^7:\left\{\ba{l}
\displaystyle{dx_7(x_1,t_1)
=\left(\alpha x_1+\tilde \beta_1\right)
h_1^{-1/2}(u_1)dx_1}
\\[3mm]
\qquad\quad\displaystyle{
+\left[\alpha
 h_1^{1/2}(u_1)
-\frac{1}{2}
\left(\alpha x_1+\tilde\beta_1\right)
h_1^{-1/2}(u_1)
\frac{dh_1}{du_1}u_{1x_1}-\lambda_3
\right]dt_1}
\\[3mm]
dt_7(x_1,t_1)=dt_1\\[3mm]
\displaystyle{h_7\left(u_7(x_7,t_7)\right)=
\left(\alpha x_1+\tilde\beta_1\right)^2},
\qquad \alpha\in\Re\backslash\{0\},\ \{\lambda_3,\tilde\beta_1\}\in\Re
\ea\right.
\ed

\bg
_2\mbox{\bf H}_3^6:\left\{\ba{l}
\displaystyle{dx_6(x_3,t_3)=
h_3^{-1/2}(u_3)dx_3
+\left[-\frac{1}{2}h_3^{-1/2}(u_3)
\frac{dh_3}{du_3}u_{3x_3}-\lambda_2x_3\right]
dt_3}
\\[3mm]
dt_6(x_3,t_3)=dt_3\\[3mm]
\displaystyle{h_6\left(u_6(x_6,t_6)\right)=\lambda_2 x_3},
\qquad \lambda_2\in\Re\backslash\{0\}.
\ea\right.
\ed

\bg
_2\mbox{\bf H}_2^8:\left\{\ba{l}
\displaystyle{dx_8(x_2,t_2)
=h_2^{-1/2}(u_2)dx_2-\left[\lambda h_2^{1/2}(u_2)
+\frac{1}{2}h_2^{-1/2}(u_2)
\frac{dh_2}{du_2}u_{2x_2}
+\lambda_8
\right]dt_2}
\\[3mm]
dt_8(x_2,t_2)=dt_2\\[3mm]
\ba[b]{c}
\scriptstyle u_8(x_8,t_8)\\
\bigint
\ea
\displaystyle{
\left\{\exp\left(\int^\xi
h_8(\eta)d\eta\right) \left[\lambda\int^\xi\exp\left(\int^{\eta}
h_8(\eta')d\eta'\right)d\eta\right]^{-1}\right\}d\xi=x_2
}
\ea\right.
\ed

\strut\hfill

Below we list the $x$-generalised hodograph transformations
$
\displaystyle{ _2\mbox{\bf H}_{\tilde n}^n}
$
by which autonomous equation $n$ is transformed in the nonautonomous
equation $\tilde n$. This refers to Cases IX -- XVI. Diagram 2 shows
the direction in which the transformations act.
\bg
_2\mbox{\bf H}_{\tilde 1}^1:\left\{\ba{l}
\displaystyle{dx_1(\tXO,\tTO)
=h_1^{1/2}(\tXO)k_1^{-1/2}(\tilde u_1)d\tXO
-\left[\frac{k_1^{1/2}(\tilde u_1) \{h_1\}_{\tXO}}{h_1^{1/2}(\tXO)}
\right.
}\\[5mm]
\qquad\quad\displaystyle{\left.
\vphantom{\frac{k_7^{1/2}(\tilde u_4)}{h_4(\tXF)}}
-\frac{1}{2}h_1^{-1/2}(\tXO)\frac{dh_1}{d\tXO}k_1^{1/2}(\tilde u_1)
+\frac{1}{2}h_1^{1/2}(\tXO)k_1^{-1/2}(\tilde u_1)
\frac{dk_1}{d\tilde u_1}\tilde u_{1\tXO}\right]d\tTO
}\\[4mm]
dt_1(\tXO,\tTO)=d\tTO\\[3mm]
\displaystyle{u_1(x_1,t_1)=\tXO}
\ea\right.
\ed

\bg
_2\mbox{\bf H}_{\tilde 2}^2:\left\{\ba{l}
\displaystyle{dx_2(\tXTW,\tTTW)
=h_2^{1/2}(\tXTW)k_2^{-1/2}(\tilde u_2)d\tXTW
-\left[\frac{k_2^{1/2}(\tilde u_2) \{h_2\}_{\tXTW}}{h_2^{1/2}(\tXTW)}
+\lambda h_2(\tXTW)
\right.
}\\[5mm]
\qquad\quad\displaystyle{\left.
\vphantom{\frac{k_7^{1/2}(\tilde u_4)}{h_4(\tXF)}}
-\frac{1}{2}h_2^{-1/2}(\tXTW)\frac{dh_2}{d\tXTW}k_2^{1/2}(\tilde u_2)
+\frac{1}{2}h_2^{1/2}(\tXTW)k_2^{-1/2}(\tilde u_2)
\frac{dk_2}{d\tilde u_2}\tilde u_{2\tXTW}\right]d\tTTW
}\\[4mm]
dt_2(\tXTW,\tTTW)=d\tTTW\\[3mm]
\displaystyle{u_2(x_2,t_2)=\tXTW}
\qquad \lambda\in\Re\backslash\{0\}
\ea\right.
\ed

\bg
_2\mbox{\bf H}_{\tilde 3}^3:\left\{\ba{l}
\displaystyle{dx_3(\tXTH,\tTTH)
=h_3^{1/2}(\tXTH)k_3^{-1/2}(\tilde u_3)d\tXTH}
\\[3mm]
\qquad\quad\displaystyle{
+\left[\left(h_3^{-1/2}(\tXTH)\frac{dh_3}{d\tXTH}
-h_3^{1/2}(\tXTH)\left(\frac{dh_3}{d\tXTH}\right)^{-1}\frac{d^2h_3}{d\tXTH^2}
\right)k_3^{1/2}(\tilde u_3)\right.
}\\[5mm]
\displaystyle{
\qquad\quad
-\frac{1}{2}h_3^{1/2}(\tXTH)k_3^{-1/2}(\tilde u_3)\frac{dk_3}{d\tilde
u_3}\tilde u_{3\tXTH}}\\ [4mm]
\displaystyle{\left.
\qquad\quad
-2\lambda_2h_3^2(\tXTH)\left(\frac{dh_3}{d\tXTH}
\right)^{-1}k_3^{-1/2}(\tilde u_3)\right]d\tTTH}\\ [3mm]
dt_3(\tXTH,\tTTH)=d\tTTH\\[3mm]
\displaystyle{u_3(x_3,t_3)=\tXTH}
\qquad \lambda_2\in\Re\backslash\{0\}
\ea\right.
\ed

\bg
_2\mbox{\bf H}_{\tilde 4}^4:\left\{\ba{l}
\displaystyle{dx_4(\tXF,\tTF)
=k_4^{-1/2}(\tilde u_4)d\tXF-\left[\frac{k_4^{1/2}(\tilde u_4)}{h_4(\tXF)}
\left(\lambda_2-\frac{dh_4}{d\tXF}\right)\right.}\\[5mm]
\qquad\quad\displaystyle{\left.
\vphantom{\frac{k_4^{1/2}(\tilde u_4)}{h_4(\tXF)}}
+\frac{1}{2}k_4^{-1/2}(\tilde u_4)\frac{dk_4}{d\tilde u_4}\tilde
u_{4\tXF}+\lambda_4+k_4^{-1/2}(\tilde u_4)h_4(\tXF)\right]d\tTF
}\\[4mm]
dt_4(\tXF,\tTF)=d\tTF\\[3mm]
\displaystyle{u_4(x_4,t_4)=\tXF},
\qquad \{\lambda_2,\lambda_4\}\in\Re
\ea\right.
\ed

\bg
_2\mbox{\bf H}_{\tilde 5}^5:\left\{\ba{l}
\displaystyle{dx_5(\tXV,\tTV)
=h_5^{1/2}(\tXV)k_5^{-1/2}(\tilde u_5)d\tXV
-\left[\frac{\{h_5\}_{\tXV}k_5^{1/2}(\tilde u_5)}{h_5^{1/2}(\tXV)}
\right.}\\[5mm]
\qquad\quad\displaystyle{\left.
\vphantom{\frac{k_4^{1/2}(\tilde u_4)}{h_4(\tXF)}}
-\frac{1}{2}h_5^{-1/2}(\tXV)\frac{dh_5}{d\tXV}k_5^{1/2}(\tilde u_5)
+\frac{1}{2}h_5^{1/2}(\tXV)k_5^{-1/2}(\tilde
u_5)\frac{dk_5}{d\tilde u_5}\tilde u_{5\tXV}\right.
}\\[4mm]
\qquad\quad\displaystyle{\left.
\vphantom{\frac{k_4^{1/2}(\tilde u_4)}{h_4(\tXF)}}
+\lambda h_5(\tXV)-\frac{\lambda_2}{\lambda}\right]d\tTV
}\\[4mm]
dt_5(\tXV,\tTV)=d\tTV\\[3mm]
\displaystyle{u_5(x_5,t_5)=\tXV},
\qquad \lambda\in \Re\backslash\{0\},\ \lambda_2\in\Re
\ea\right.
\ed

\bg
_2\mbox{\bf H}_{\tilde 6}^6:\left\{\ba{l}
\displaystyle{dx_6(\tXS,\tTS)
=k_6^{-1/2}(\tilde u_6)d\tXS-
\left[k_6^{1/2}(\tilde u_6)\left(\frac{d h_6}{d\tXS}\right)^{-1}
\frac{d^2h_6}{d\tXS^2}\right.
}\\[5mm]
\qquad\quad\displaystyle{\left.
\vphantom{k_6^{1/2}(\tilde u_6)\left(\frac{d h_6}{d\tXS}\right)^{-1}
\frac{d^2h_6}{d\tXS^2}}
+\frac{1}{2}k_6^{-1/2}(\tilde u_6)\frac{dk_6}{d\tilde u_6} \tilde
u_{6\tXS}
+h_6(\tXS)\right]d\tTS}\\[4mm]
dt_6(\tXS,\tTS)=d\tTS\\[3mm]
\displaystyle{u_6(x_6,t_6)=\tXS}
\ea\right.
\ed

\bg
_2\mbox{\bf H}_{\tilde 7}^7:\left\{\ba{l}
\displaystyle{dx_7(\tXSV,\tTSV)
=h_7^{1/2}(\tXSV)k_7^{-1/2}(\tilde u_7)d\tXSV
-\left[\frac{k_7^{1/2}(\tilde u_7) \{h_7\}_{\tXSV}}{h_7^{1/2}(\tXSV)
}+\lambda_3
\right.
}\\[5mm]
\qquad\quad\displaystyle{\left.
\vphantom{\frac{k_7^{1/2}(\tilde u_4)}{h_4(\tXF)}}
-\frac{1}{2}h_7^{-1/2}(\tXSV)\frac{dh_7}{d\tXSV}k_7^{1/2}(\tilde u_7)
+\frac{1}{2}h_7^{1/2}(\tXSV)k_7^{-1/2}(\tilde u_7)
\frac{dk_7}{d\tilde u_7}\tilde u_{7\tXSV}\right]d\tTSV
}\\[4mm]
dt_7(\tXSV,\tTSV)=d\tTSV\\[3mm]
\displaystyle{u_7(x_7,t_7)=\tXSV},
\qquad \lambda_3\in\Re
\ea\right.
\ed

\bg
_2\mbox{\bf H}_{\tilde 8}^8:\left\{\ba{l}
\displaystyle{dx_8(\tXE,\tTE)
=k_8^{-1/2}(\tilde u_8)d\tXE}\\[5mm]
\qquad\quad\displaystyle{
-\left[h_8(\tXE) k_8^{1/2}(\tilde u_8)
+\frac{1}{2}k_8^{-1/2}(\tilde u_8)\frac{dk_8}{d\tilde u_8}\tilde
u_{8\tXE}
+\lambda_8\right]d\tTE}\\[4mm]
dt_8(\tXE,\tTE)=d\tTE\\[3mm]
\displaystyle{u_8(x_8,t_8)=\tXE}
\qquad \lambda_8\in\Re
\ea\right.
\ed

\strut\hfill

\strut\hfill



\label{euler-lastpage}
\end{document}